\documentstyle[12pt,epsf]{article}
\begin{document}
\def\<{\langle}\def\>{\rangle}
\def\cj#1{\overline{#1}}
\title{Homodyning as universal detection\footnote{To appear in 
{\em Quantum Communication and Measurement}, ed. 
V. P. Belavkin, O. Hirota, and R. L. Hudson, Plenum Press 
(New York and London 1997)}}
\author{Giacomo Mauro D'Ariano\\
{\em Department of Electrical and Computer Engineering},\\
{\em Department of Physics and Astronomy},\\
{\em Northwestern University, Evanston, IL  60208}\\
and\\
{\em Istituto Nazionale di Fisica Nucleare, Sezione di Pavia,}\\
{\em via A. Bassi 6, I-27100 Pavia, Italy}}
\date{}
\maketitle
\begin{abstract}
Homodyne tomography---i. e. homodyning while scanning the local oscillator 
phase---is now a well assessed method for ``measuring'' the quantum
state. In this paper I will show how it can be used as a kind of 
universal detection, for measuring generic field operators, 
however at expense of some additional noise. The general class of 
field operators that can be measured in this way is presented, and 
includes also operators that are inaccessible to heterodyne 
detection. The noise from tomographical homodyning is compared to that
from heterodyning, for those operators that can be measured in both
ways. It turns out that for some operators homodyning is better than
heterodyning when the mean photon number is sufficiently small. 
Finally, the robustness of the method to additive phase-insensitive
noise is analyzed. It is shown that just half photon
of thermal noise would spoil the measurement completely.
\end{abstract}
\small
\section{Introduction}
Homodyne tomography is the only viable method currently known for determining 
the detailed state of a quantum harmonic oscillator---a mode of the
electromagnetic field.
The state measurement is achieved by repeating many homodyne 
measurements at different phases $\phi$ with respect to the local 
oscillator (LO). The experimental work of the group in 
Eugene-Oregon \cite{Raymer1}
undoubtedly established the feasibility of the method, even though the
earlier data analysis were based on a filtered procedure that affected the
results with systematic errors. Later, the theoretical group in
Pavia-Italy presented an exact reconstruction algorithm \cite{our}, which is
the method currently adopted in actual experiments (see, for
example, Refs. \cite{munroe1} and \cite{schiller}). 
The reconstruction algorithm of 
Ref. \cite{our} was later greatly simplified
\cite{dlp}, so that it was possible also to recognize the
feasibility of the method even for nonideal quantum efficiency $\eta<1$
at the homodyne detector, and, at the same time, establishing lower bounds for $\eta$ 
for any given matrix representation. 
After these first results, further theoretical progress has been made,
understanding the mechanisms that underly the generation of statistical 
errors \cite{dariJEOS}, thus limiting the sensitivity of the method. 
More recently, for $\eta=1$ non trivial factorization formulas have been 
recognized \cite{Richter,factulf} for the 
``pattern functions'' \cite{lpd} that are necessary to reconstruct 
the photon statistics.
\par In this paper I will show how homodyne tomography can also be used
as a method for measuring generic field operators. In fact, due to 
statistical errors, the measured matrix elements cannot 
be used to obtain expectations of field operators, and a
different algorithm for analyzing homodyne data is needed
suited to the particular field operator whose expectation one wants to estimate.
Here, I will present an algorithm valid for any operator that admits a normal ordered
expansion, giving the general class of operators that can be measured
in this way, also as a function of the quantum efficiency
$\eta$. Hence, from the same bunch of homodyne experimental data, now
one can obtain not only the density matrix of the state, but also
the expectation value of various field operators, including some 
operators that are inaccessible to heterodyne 
detection. However, the price to pay for such detection flexibility
is that all measured quantities will be affected by noise. 
But, if one compares this noise with that from heterodyning 
(for those operators that can be measured in both
ways), it turns out that for some operators homodyning is less noisy
than heterodyning, at least for small mean photon numbers.
\par Finally, I will show that the method of homodyne tomography is quite robust
to sources of additive noise. Focusing attention on the most common situation 
in which the noise is Gaussian and independent on the LO phase, I will
show that this kind of noise produces the same effect of nonunit
quantum efficiency at detectors. Generalizing the result of 
Ref. \cite{dlp}, I will give bounds for the overall rms noise 
level below which the tomographical reconstruction is still possible.
I will show that the smearing effect of half photon of thermal noise 
in average is sufficient to completely spoil the measurement, making
the experimental errors growing up unbounded.
\section{Short up-to-date review on homodyne tomography}
The homodyne tomography method is designed to obtain a general matrix element
$\langle\psi|\hat{\varrho}|\varphi\rangle$ in form of expectation of a
function of the homodyne outcomes at different phases with
respect to the LO. In equations, one has
\begin{eqnarray}
\langle\psi|\hat{\varrho}|\varphi\rangle=\int^{\pi}_0\!\frac{d\phi }{\pi}\; 
\int^{+\infty}_{-\infty}\!\! d x\,p(x;\phi )\;f_{\psi\varphi}(x;\phi)\;,
\label{rad1c}
\end{eqnarray}
where $p(x;\phi)$ denotes the probability distribution
of the outcome $x$ of the quadrature $\hat x_{\phi}=\frac{1}{2}
\left(a^{\dagger} e^{i\phi}+a e^{-i\phi}\right)$ of the field mode
with particle operators $a$ and $a^{\dagger}$ at phase $\phi$ with
respect to the LO. Notice that it is sufficient to average only over 
$\phi\in [0,\pi]$, due to the symmetry $\hat x_{\phi+\pi}=-\hat x_{\phi}$.
One wants the function $f_{\psi\varphi}(x;\phi)$ bounded for all $x$,
whence every moment will be bounded for any possible ({\em a priori} unknown)
probability distribution $p(x;\phi)$. Then, according to the
central-limit theorem, one is guaranteed that the integral 
in Eq. (\ref{rad1c}) can be sampled statistically over a 
sufficiently large set of data, and the average values for 
different experiments will be Gaussian distributed, allowing estimation of
confidence intervals. If, on the other hand, the kernel $f_{\psi\varphi}(x;\phi)$ 
turns out to be unbounded, then we will say that the matrix element cannot be measured
by homodyne tomography.
\par The easiest way to obtain the integral kernel $f_{\psi\varphi}(x;\phi)$
is starting from the operator identity
\begin{equation}
\hat{\varrho} = \int \frac{d^2\alpha}{\pi}\,\hbox{Tr}
(\hat{\varrho} e^{-\overline{\alpha} a+\alpha
a^{\dagger}}) \, e^{-\alpha a^{\dagger}+
\overline{\alpha}a}\label{ooop}
\end{equation}
which, by changing to polar variables $\alpha = (i/2)k
e^{i\phi}$, becomes
\begin{equation}
\hat{\varrho} = \int^{\pi}_0 \, \frac{d\phi}{\pi}\, \int^{+\infty}_{-\infty}\,
\frac{d k\, |k|}{4}\,\hbox{Tr}
(\hat{\varrho} e^{ik\hat{x}_{\phi}})\, e^{-ik\hat{x}_{\phi}}\, .
\label{op}
\end{equation}
Equation (\ref{ooop}) is nothing but the operator form of the
Fourier-transform relation between Wigner function and 
characteristic function: it can also be considered as an operator form of
the Moyal identity
\begin{eqnarray}
\int\frac{d^2 z}{\pi}\langle k|\hat D^{\dag}(z)|m\rangle \langle l|\hat D(z)|
n\rangle &=&\langle k|n\rangle \langle l|m\rangle \;.
\end{eqnarray}
The trace-average in Eq. (\ref{op}) can be evaluated in
terms of $p(x,{\phi})$, using the complete set
$\{|x\rangle_{\phi}\}$ of eigenvectors of $\hat{x}_{\phi}$, and
exchanging the integrals over $x$ and $k$. One obtains
\begin{equation}
\hat{\varrho} = \int^{\pi}_0 \frac{d\phi}{\pi}
\int^{+\infty}_{-\infty} \, d x\, p(x;\phi) K(x-
\hat{x}_{\phi}) \, ,
\label{p1}
\end{equation}
where the integral kernel $K(x)$ is given by 
\begin{equation}
K(x) = -\frac{1}{2} \hbox{P} \frac{1}{x^2} \equiv
-\lim_{\varepsilon \rightarrow 0^+} \frac{1}{2} \hbox{Re}
\frac{1}{(x+i\varepsilon)^2}\,,
\label{kf}
\end{equation}
P denoting the Cauchy principal value.
Taking matrix elements of both sides of Eq. (\ref{p1}) between vectors
$\psi$ and $\varphi$, we obtain the sampling formula we were looking
for, namely
\begin{equation}
\langle\psi |\hat{\varrho}|\varphi\rangle = \int^{\pi}_0 
\frac{d\phi}{\pi}
\int^{+\infty}_{-\infty} \, d x\, p(x;\phi)\langle\psi | K(x-
\hat{x}_{\phi})|\varphi\rangle\,.\label{samp1}
\end{equation}
Hence, the matrix element $\langle\psi |\hat{\varrho}|\varphi\rangle$
is obtained by averaging the function $f_{\psi\varphi}(x;\phi)\equiv
\langle\psi | K(x-\hat{x}_{\phi})|\varphi\rangle$ over homodyne data
at different phases $\phi$.
As we will see soon, despite $K(x)$ is unbounded, for particular vectors 
$\psi$ and $\varphi$ in the Hilbert space the matrix element 
$\langle \psi |K(x-\hat{x}_{\phi})|\varphi\rangle$ is bounded, and
thus the integral (\ref{samp1}) can be sampled experimentally.
\par Before analyzing specific matrix representations, I recall how 
the sampling formula (\ref{samp1}) can be generalized to the case of 
nonunit quantum efficiency. Low efficiency homodyne detection simply produces a
probability $p_{\eta}(x;\phi)$ that is a Gaussian convolution of 
the ideal probability $p(x;\phi)$ for $\eta=1$ (see, for example, Ref.
\cite{Bilk-poms}). In terms of the generating functions of the 
$\hat{x}_{\phi}$-moments one has 
\begin{equation}
\int^{+\infty}_{-\infty} \! d x\, p_{\eta} (x;\phi)e^{ik
x} = \exp\left( -\frac{1-\eta}{8\eta}k^2\right) \, 
\int^{+\infty}_{-\infty}\! d x\, p(x;\phi)e^{ik x} \, .\label{gc}
\end{equation}
Upon substituting Eq. (\ref{gc}) into Eq. (\ref{op}),
and by following the same lines that lead us to Eq. (\ref{p1}),
one obtains the operator identity
\begin{equation}
\hat{\varrho} = \int^{\pi}_0 \frac{d\phi}{\pi} \,
\int^{+\infty}_{-\infty} \, d x\, p_{\eta} (x;\phi)
K_{\eta} (x-\hat{x}_{\phi}) \, ,
\label{p2}
\end{equation}
where now the kernel reads
\begin{equation}
K_{\eta} (x) = \frac{1}{2} \hbox{Re} \int^{+\infty}_0 \, d k\,
k\,\exp \left( \frac{1-\eta}{8\eta}k^2+ikx\right) \, .\label{kf2}
\end{equation}
The desired sampling formula for 
$\langle\psi |\hat{\varrho}|\varphi\rangle$ is obtained again as in
Eq. (\ref{samp1}), by taking matrix elements of both sides of Eq. (\ref{kf2}). Notice
that now the kernel $K_{\eta} (x)$ is not even a tempered
distribution: however, as we will see immediately, the matrix elements
of $K_{\eta} (x-\hat x_{\phi})$ are bounded for 
some representations, depending on the value of $\eta$.
The matrix elements $\langle\psi|K_{\eta} (x-\hat a_{\phi})|\varphi\rangle$
are bounded if the following inequality is satisfied for all phases $\phi\in[0,\pi]$
\begin{equation}
\eta>\frac{1}{1+4\varepsilon^2(\phi)}\, ,
\label{eta1}
\end{equation}
where $\varepsilon^2(\phi)$ is the harmonic mean
\begin{equation}
\frac{2}{\varepsilon^2(\phi)} =
\frac{1}{\varepsilon^2_{\psi}(\phi)} +
\frac{1}{\varepsilon^2_{\varphi}(\phi)}\;,
\end{equation}
and $\varepsilon^2_{\upsilon}(\phi)$ is the ``resolution'' of
the vector $|\upsilon\rangle$ in the
$\hat{x}_{\phi}$-representation, namely:
\begin{equation}
|{}_{\phi}\langle x|\upsilon\rangle|^2\simeq \exp \left[ -
\frac{x^2}{2\varepsilon^2_{\upsilon}(\phi)}\right]  \,.
\label{lead}
\end{equation}
In Eq. (\ref{lead}) the symbol $\simeq$ stands for the
leading term as a function of $x$, and $|x\rangle_{\phi} \equiv
e^{i a^{\dagger} a\phi}|x\rangle$ denote eigen-ket of the
quadrature $\hat{x}_{\phi}$ for eigenvalue $x$. Upon maximizing Eq.
(\ref{eta1}) with respect to $\phi$ one obtains the bound
\begin{equation}
\eta > \frac{1}{1+4\varepsilon^2}\;, \qquad \varepsilon^2
=\, \min_{\phi\in[0,\pi]} \{\varepsilon^2(\phi)\} \,.
\label{eta2}
\end{equation}
One can easily see that the bound is $\eta > 1/2$ for both number-state
and coherent-state representations, whereas it is $\eta>(1+s^2)^{-1}\ge 1/2$
for squeezed-state representations with minimum squeezing factor 
$s<1$. On the other hand, for the quadrature representation
one has $\eta >1$, which means that this matrix representation cannot
be measured. The value $\eta = 1/2$ is actually an absolute bound for 
all representations satisfying the ``Heisenberg relation''
$\epsilon(\phi)\epsilon(\phi+\frac{\pi}{2})\geq \frac{1}{4}$ with the
equal sign, which include all known representations 
(for a discussion on the existence of
exotic representations see Ref. \cite{Bilk-tomo}). Here, I want to 
emphasize that the existence of such a lower bound for quantum efficiency 
is actually of fundamental relevance, as it prevents measuring the wave
function of a single system using schemes of weak repeated indirect 
measurements on the same system \cite{single}.
\par At the end of this section, from Ref. \cite{dlp} I report for 
completeness the kernel $\<n|K(x-\hat x_{\phi})|m\>$ for matrix elements 
between number eigenstates. One has
\begin{eqnarray}
&&\langle n|K_{\eta} (x-\hat{x}_{\phi})|n+d\rangle =e^{-id \phi} 
2\kappa^{d +2}\sqrt{\frac{n!}{(n+d)!}}e^{-\kappa^2x^2}\label{numb}\\
&\times&\sum^n_{\nu=0} \, \frac{(-)^{\nu}}{\nu!} \left({n+d
\atop n-\nu}\right) (2\nu+d+1)!\kappa^{2\nu}
\hbox{Re} \,\left\{ (-i)^{d} D_{-
(2\nu+d+2)} (-2i\kappa x)\, \right\}\;,\nonumber
\end{eqnarray}
where $\kappa =\sqrt{\eta/(2\eta-1)}$, and 
$D_{\sigma}(z)$ denotes the parabolic cylinder function. For $\eta=1$
the kernel factorizes as follows \cite{Richter,factulf}
\begin{eqnarray}
\langle n|K(x-\hat x_{\phi})|n+d\rangle\nonumber\\
= e^{-id \phi}[2xu_n(x)v_{n+d}(x)-\sqrt{n+1}u_{n+1}(x)v_{n+d}(x)-
\sqrt{m+1}u_n(x)v_{n+d+1}(x)]\;,
\label{fact1}
\end{eqnarray}
where $u_n(x)$ and $v_n(x)$ are the regular and irregular energy eigen-functions of 
the harmonic oscillator
\begin{eqnarray}
u_j(x)=\frac{1}{\sqrt{j!}}
\left(x-\frac{\partial_x}{2}\right)^j\left(\frac{2}{\pi}\right)^{1/4}e^{-x^2}
,\qquad v_j(x)=\frac{1}{\sqrt{j!}}
\left( x-\frac{\partial_x}{2}\right)^j \left(2\pi\right)^{1/4} e^{-x^2}
\int_0^{\sqrt{2}x} \mbox{d}t \, e^{t^2}. 
\end{eqnarray}
\section{Measuring generic field operators}
Homodyne tomography provides the maximum achievable information 
on the quantum state, and, in principle, the knowledge of the density 
matrix should allow one to calculate the expectation value $\<\hat
O\>=\mbox{Tr}[\hat O\hat\varrho]$ of any observable $\hat O$. 
However, this is generally true only when one has an analytic
knowledge of the density matrix, but it is not true when the matrix
has been obtained experimentally. In fact, the Hilbert space is
actually infinite dimensional, whereas experimentally one can
achieve only a finite matrix, each element being affected by
an experimental error. Notice that, even though the method allows
one to extract {\em any} matrix element in the Hilbert space 
from the same bunch of experimental data, however, it is 
the way in which errors converge in the Hilbert space 
that determines the actual possibility of estimating the trace 
$\mbox{Tr}[\hat O\hat\varrho]$. To make things more concrete, let 
us fix the case of the number representation, and suppose we want 
to estimate the average photon number $\<a^{\dag}a\>$. 
In Ref. \cite{nico} it has been shown that for nonunit quantum 
efficiency the statistical error for the diagonal matrix element 
$\< n|\hat\varrho|n\>$ diverges faster than exponentially
versus $n$, whereas for $\eta=1$ the error saturates for large $n$ to the
universal value $\varepsilon_n=\sqrt{2/N}$ that depends only on 
the number $N$ of experimental data, but is independent on both $n$ and on 
the quantum state. Even for the unrealistic case $\eta=1$, one can 
see immediately that the estimated expectation value
$\<a^{\dag}a\>=\sum_{n=0}^{H-1}n\varrho_{nn}$ 
based on the measured matrix elements $\varrho_{nn}$, is not guaranteed 
to converge versus the truncated-space dimension $H$, because
the error on $\varrho_{nn}$ is nonvanishing versus $n$.
Clearly in this way I am not proving that the expectation
$\<a^{\dag}a\>$ is unobtainable from homodyne data, because
matrix errors convergence depends on the chosen representation basis, whence the
ineffectiveness of the method may rely in the data processing,
more than in the actual information contained in the bunch of experimental data.
Therefore, the question is: is it possible to estimate a generic expectation value 
$\<\hat O\>$ directly from homodyne data, without using the measured density matrix?
As we will see soon, the answer is positive in most
cases of interest, and the procedure for estimating the expectation $\<\hat O\>$
will be referred to as {\em homodyning the observable} $\hat O$. 
\par By {\em homodyning the observable} $\hat O$ I mean averaging an appropriate 
kernel function ${\cal R}[\hat O](x;\phi)$ (independent on the state
$\hat\varrho$) over the experimental homodyne data,
achieving in this way the expectation value of the observable $\<\hat O\>$
for every state $\hat\varrho$. Hence, the kernel
function ${\cal R}[\hat O](x;\phi)$ is defined through the identity
\begin{equation}
\langle\hat O\rangle = \int^{\pi}_0 \frac{d\phi}{\pi}
\int^{+\infty}_{-\infty} \, d x\, p(x;\phi) {\cal R}[\hat O](x;\phi)
\,.\label{sampO}
\end{equation}
From the definition of ${\cal R}[\hat O](x;\phi)$ in Eq. (\ref{sampO}), 
and from Eqs. (\ref{ooop}) and (\ref{op})---which generally hold true 
for any Hilbert-Schmidt operator in place of $\hat\varrho$---one obtains 
\begin{eqnarray}
\hat O=\int^{\pi}_0 \frac{d\phi}{\pi}
\int^{+\infty}_{-\infty} \, d x\, {\cal R}[\hat O](x;\phi)
|x\rangle_{\phi} {}_{\phi}\langle x|\;,
\end{eqnarray}
with the kernel ${\cal R}[\hat O](x;\phi)$ given by
\begin{eqnarray}
{\cal R}[\hat O](x;\phi)=\mbox{Tr}[\hat O K(x-\hat x_{\phi})]\;,\label{caspita}
\end{eqnarray}
and $K(x)$ given in Eq. (\ref{kf}).
The validity of Eq. (\ref{caspita}), however, is limited only to the case 
of a Hilbert-Schmidt operator $\hat O$, otherwise it is ill defined.
Nevertheless, one can obtain the explicit form
of the kernel ${\cal R}[\hat O](x;\phi)$ in a different way. 
Starting from the identity involving trilinear products of Hermite 
polynomials \cite{gradshtein}
\begin{eqnarray}
\int^{+\infty}_{-\infty}\, d x\,e^{-x^2}\,H_k(x)\,H_m(x)\,H_n(x)=\frac{
2^{\frac{m+n+k}{2}}\pi^{{1\over2}}k!m!n!}{(s-k)!(s-m)!(s-n)!}\;,
\quad\mbox{for }k+m+n=2s\mbox{ even}\;,
\end{eqnarray}
Richter proved the following nontrivial formula for the
expectation value of the normally ordered field operators \cite{Rich}
\begin{eqnarray}
\<a^{\dag}{}^n a^m\>=\int^{\pi}_0 \frac{d\phi}{\pi}
\int^{+\infty}_{-\infty} \, d x\, p(x;\phi) e^{i(m-n)\phi}\frac{H_{n+m}(\sqrt{2}x)}{
\sqrt{2^{n+m}}{{n+m}\choose n}}\;,
\end{eqnarray}
which corresponds to the kernel
\begin{eqnarray}
{\cal R}[a^{\dag}{}^n a^m](x;\phi)=e^{i(m-n)\phi}
\frac{H_{n+m}(\sqrt{2}x)}{\sqrt{2^{n+m}}{{n+m}\choose n}}\;.
\end{eqnarray}
This result can be easily extended to the case of nonunit quantum 
efficiency $\eta< 1$, as
the normally ordered expectation $\<a^{\dag}{}^n
a^m\>$ just gets an extra factor $\eta^{{1\over2}(n+m)}$. Therefore, one has
\begin{eqnarray}
{\cal R}_{\eta}[a^{\dag}{}^n a^m](x;\phi)=e^{i(m-n)\phi}
\frac{H_{n+m}(\sqrt{2}x)}{\sqrt{(2\eta)^{n+m}}{{n+m}\choose n}}\;,\label{Rnm}
\end{eqnarray}
where the kernel ${\cal R}_{\eta}[\hat O](x;\phi)$ is defined as in
Eq. (\ref{sampO}), but with the experimental probability distribution 
$p_{\eta}(x;\phi)$. From Eq. (\ref{Rnm}) by linearity on can obtain
the kernel ${\cal R}_{\eta}[\hat f](x;\phi)$ for any operator function
$\hat f$ that has normal ordered expansion
\begin{eqnarray}
\hat f\equiv f(a,a^{\dag})=\sum_{nm=0}^{\infty}f^{(n)}_{nm}a^{\dag}{}^n a^m\;.
\end{eqnarray}
From Eq. (\ref{Rnm}) one obtains
\begin{eqnarray}
{\cal R}_{\eta}[\hat f](x;\phi)=\sum_{s=0}^{\infty}\frac{H_s(\sqrt{2}x)}{s!
(2\eta)^{s/2}}\sum_{nm=0}^{\infty}f^{(n)}_{nm}e^{i(m-n)\phi}n!m!\delta_{n+m,s}=
\sum_{s=0}^{\infty}\frac{H_s(\sqrt{2}x)i^s}{s!
(2\eta)^{s/2}}\frac{d^s}{dv^s}\Bigg|_{v=0}\!\!\!\!
{\cal F}[\hat f](v;\phi),\label{123}
\end{eqnarray}
where
\begin{eqnarray}
{\cal F}[\hat f](v;\phi)=\sum_{nm=0}^{\infty}f^{(n)}_{nm}
{{n+m}\choose m}^{-1}(-iv)^{n+m}e^{i(m-n)\phi}\;.\label{ff}
\end{eqnarray}
Continuing from Eq. (\ref{123}) one obtains
\begin{eqnarray}
{\cal R}_{\eta}[\hat f](x;\phi)=\exp\left(\frac{1}{2\eta}\frac{d^2}{dv^2}+
\frac{2ix}{\sqrt{\eta}}\frac{d}{dv}\right)\Bigg|_{v=0}{\cal F}[\hat f](v;\phi)\;,
\end{eqnarray}
and finally  
\begin{eqnarray}
{\cal R}_{\eta}[\hat f](x;\phi)=\int_{-\infty}^{+\infty}\frac{dw}{\sqrt{2\pi\eta^{-1}}}
e^{-\frac{\eta}{2}w^2}{\cal F}[\hat f](w+2ix/\sqrt{\eta};\phi)\;.\label{generalf}
\end{eqnarray}
Hence one concludes that the operator $\hat f$ can be measured by
homodyne tomography if the function ${\cal F}[\hat f](v;\phi)$ in 
Eq. (\ref{ff}) grows slower than $\exp (-\eta v^2/2)$ for
$v\to\infty$, and the integral in Eq. (\ref{generalf}) grows at most
exponentially for $x\to\infty$ (assuming $p(x;\phi)$ goes to zero
faster than exponentially at $x\to\infty$).
\par In Table \ref{td1} I report the kernel ${\cal R}_{\eta}[\hat O](x;\phi)$
for some operators $\hat O$. One can see that for the raising
operator $\hat e_+$ the kernel diverges at $\eta=1/2^+$, namely it
can be measured only for $\eta>1/2$. The operator
$\hat W_s$ in the same table gives the generalized Wigner function 
$W_s (\alpha,\bar{\alpha})$ for ordering parameter $s$ through the
identity $W_s (\alpha,\bar{\alpha})=\mbox{Tr}[\hat D(\alpha)\hat\varrho
\hat D^{\dag}(\alpha)\hat W_s]$. From the expression of 
${\cal R}_{\eta}[\hat W_s](x;\phi)$ it follows that
by homodyning with quantum efficiency $\eta$ one can measure the
generalized Wigner function only for $s<1-\eta^{-1}$: in particular, as
already noticed in Refs. \cite{dlp}, the usual Wigner function for $s=0$
cannot be measured for any quantum efficiency [in fact one would have
${\cal R}_{1}[\hat D^{\dag}(\alpha)\hat W_0\hat D(\alpha)](x;\phi)=
K [x-\mbox{Re}(\alpha e^{-i\phi})]$, with $K (x)$
unbounded as given in Eq. (\ref{kf})].
\begin{table}[hbt]
\begin{center}
\vspace{-20truept}
\begin{tabular}{|c|c|c|}
\hline \hline
& $\hat O$ & ${\cal R}_{\eta}[\hat O](x;\phi)$ \\
\hline \hline 
(1) & $a^{\dag}{}^n a^m$ & ${\displaystyle e^{i(m-n)\phi}
\frac{H_{n+m}(\sqrt{2}x)}{\sqrt{2^{n+m}}{{n+m}\choose n}} }$ \\ \hline
(2) & $a$ & $2e^{i\phi}x$ \\ \hline
(3) & $a^2$ & $e^{2i\phi}(4x^2-1)$ \\ \hline
(4) & $a^{\dag}a$ & $2x^2-{1\over2}$ \\ \hline
(5) & $(a^{\dag}a)^2$ & ${8\over3}x^4-2x^2$ \\ \hline
(6) & $:\hat D^{\dag}(\alpha):\doteq e^{-\alpha a^{\dag}}e^{\cj{\alpha}a}$ 
& ${\displaystyle 
\frac{\exp[-\frac{1}{2\eta}(\cj{\alpha}e^{i\phi})^2+\frac{2x}{\sqrt{\eta}}
\cj{\alpha}e^{i\phi}]}{1+\frac{\alpha}{\cj{\alpha}}e^{-2i\phi}}+
\frac{\exp[-\frac{1}{2\eta}(\alpha e^{-i\phi})^2-\frac{2x}{\sqrt{\eta}}
\alpha e^{-i\phi}]}{1+\frac{\cj{\alpha}}{\alpha} e^{2i\phi}}}$ \\ \hline
(7) & $\hat e_{+}\doteq a^{\dag}\frac{1}{\sqrt{1+a^{\dag}a}}$ & 
${\displaystyle 2xe^{-i\phi}\frac{1}{\sqrt{2\pi\eta}}\int_{-\infty}^{+\infty}dv\,
\frac{e^{-v^2}}{(1+z)^2}\Phi\left(2,{3\over2};\frac{x^2}{1+z^{-1}}\right)}\,,\quad
z=\frac{e^{-v^2}-1}{2\eta}$ \\ \hline
(8) & $\hat W_s\doteq\frac{2}{\pi(1-s)}\left(\frac{s+1}{s-1}\right)^{a^{\dag}a}$ & 
${\displaystyle \int_0^{\infty}dt\frac{2e^{-t}}{\pi(1-s)-{1\over\eta}}
\cos\left(2\sqrt{\frac{2t}{(1-s)-{1\over\eta}}}x\right)}$ \\ \hline
(9) & $|n+d\>\< n|$ & $\< n|K(x-\hat x_{\phi})|n+d\>$ in Eqs. (\protect\ref{numb}) and
(\protect\ref{fact1}) \\
\hline \hline
\end{tabular} 
\caption{\footnotesize Kernel ${\cal R}_{\eta}[\hat O](x;\phi)$, as defined in Eq. 
(\protect\ref{sampO}), for some operators $\hat O$. [The symbol 
$\Phi(a,b;x)$ denotes the customary confluent hypergeometric 
function.]}\label{td1}
\vspace{-10truept}
\end{center}
\end{table}
\subsection{Comparison between homodyne tomography and heterodyning}
We have seen that from the same bunch of homodyne tomography data, 
not only one can recover the density matrix of the field, but also 
one can measure any field observable 
$\hat f\equiv f(a,a^{\dag})$ having {\em normal ordered} 
expansion $\hat f\equiv f^{(n)}(a,a^{\dag})=
\sum_{nm=0}^{\infty}f^{(n)}_{nm}a^{\dag}{}^n a^m$ and
bounded integral in Eq. (\ref{generalf})---this holds true 
in particular for any polynomial function of the annihilation and 
creation operators. This situation can be compared with the 
case of heterodyne detection, where again one measures general
field observables, but admitting {\em anti-normal ordered}
expansion $\hat f\equiv f^{(a)}(a,a^{\dag})=
\sum_{nm=0}^{\infty}f^{(a)}_{nm}a^ma^{\dag}{}^n$, in which case the
expectation value is obtained through the heterodyne average
\begin{eqnarray}
\<\hat f\>=\int {{d^2\alpha}\over\pi} f^{(a)}(\alpha,\cj{\alpha})
\<\alpha|\hat\varrho|\alpha\>\;.\label{QI}
\end{eqnarray}
For $\eta=1$ the heterodyne probability is just the $Q$-function 
$Q(\alpha,\cj{\alpha})={1\over\pi}\<\alpha|\hat\varrho|\alpha\>$,
whereas for $\eta=1$ it will be Gaussian convoluted.
As shown by Baltin \cite{baltin}, generally the anti-normal expansion
either is not defined, or is {\em not consistent} on the Fock basis, namely
$f^{(a)}(a,a^{\dag})|n\>$ has infinite norm or is different from
$\hat f(a,a^{\dag})|n\>$ for some
$n\geq 0$. In particular, let us focus attention on functions of the
number operator $f(a^{\dag}a)=\sum_{l=0}^{\infty}c_l(a^{\dag}a)^l$, 
$f^{(n)}(a^{\dag}a)=\sum_{l=0}^{\infty}c^{(n)}_l a^{\dag}{}^l a^l$, 
$f^{(a)}(a^{\dag}a)=\sum_{l=0}^{\infty}c^{(a)}_l a^l a^{\dag}{}^l$.
Baltin has shown that \cite{baltin}
\begin{eqnarray}
c^{(n)}_l&=&\frac{1}{l!}\int_{-\infty}^{+\infty}d\lambda\, 
g(\lambda)(e^{-i\lambda}-1)^l=\sum_{k=0}^l\frac{(-)^{l-k}f(k)}{k!(l-k)!}\;,
\nonumber\\
c^{(a)}_l&=&\frac{1}{l!}\int_{-\infty}^{+\infty}d\lambda\,e^{i\lambda} 
g(\lambda)(1-e^{i\lambda})^l=\sum_{k=0}^l\frac{(-)^k
f(-k-1)}{k!(l-k)!}\;, \label{eqBaltin}\\
g(\lambda)&\doteq&\int_{-\infty}^{+\infty}\frac{dx}{2\pi}f(x)e^{i\lambda x}
\;.\nonumber
\end{eqnarray}
From Eqs. (\ref{eqBaltin}) one can see that the normal ordered 
expansion is always well defined, whereas the anti-normal ordering 
needs extending the domain of $f$ to negative integers. However,
even though the anti-normal expansion is defined, this does not mean
that the expectation of $f(a^{\dag}a)$ can be obtained through
heterodyning, because the integral in
Eq. (\ref{QI}) may not exist. Actually, this is the case when the
anti-normal expansion is not consistent on the Fock basis. In fact,
for the exponential function $f(a^{\dag}a)=\exp(-\mu a^{\dag}a)$
one has $f^{(a)}(|\alpha|^2)=e^{\mu}\exp[(1-e^{\mu})|\alpha|^2]$; on
the Fock basis $f^{(a)}(a^{\dag}a)|n\>$ is a binomial expansion
with finite convergence radius, and this gives the consistency
condition $|1-e^{\mu}|<1$. However, one can take the analytic
continuation corresponding for $1-e^{\mu}<1$,
which coincides with the condition that the
integral in Eq. (\ref{QI}) exists for any state $\hat\varrho$ 
(the $Q$-function vanishes as $\exp(-|\alpha|^2)$ for $\alpha\to\infty$, 
at least for states with limited photon number). This argument can be
extended by Fourier transform to more general functions $f(a^{\dag}a)$, 
leading to the conclusion that there are field operators that cannot
be heterodyne-measured, even though they have well
defined anti-normal expansion, but the expansion is not consistent on the Fock basis.
As two examples, I consider the field operators $\hat e_+$ and
$\hat W_s$ in Table \ref{td1}. According to Eqs. (\ref{eqBaltin}) 
it follows that the operator $\hat e_+$ does not admit an anti-normal
expansion, whence it cannot be heterodyne detected. This is in agreement
with the fact that according to Table \ref{td1} we can homodyne $\hat e_+$
only for $\eta>1/2$, and heterodyning is equivalent to homodyning with effective
quantum efficiency $\eta=1/2$ (which corresponds to the 3 dB
noise due to the joint measurement \cite{yuen}). The case of the
operator $\hat W_s$ is different. It admits both normal-ordered and
anti-normal-ordered forms: $\hat W_s=\frac{2}{\pi(1-s)}
:\exp\left(-\frac{2}{1-s}a^{\dag}a\right):=-\frac{2}{\pi(1+s)}
:\exp\left(\frac{2}{1+s}a^{\dag}a\right):_A$, where $:\ldots:$ denotes normal ordering
and $:\ldots:_A$ anti-normal. However, the consistency condition for 
anti-normal ordering is $2/(s+1)<1$, with $s\le 1$, which
implies that one can heterodyne $\hat W_s$ for $s>-1$, again in
agreement with the value of $s$ achievable by homodyne tomography at $\eta=1/2$. 
\par Now I briefly analyze the additional noise from homodyning field
operators, and compare them with the heterodyne noise.
For a complex random variable $z=u+iv$ the noise is given by the eigenvalues
$N^{(\pm)}=\overline{|z|^2-|\overline{z}|^2}\pm|\overline{z^2}-\overline{z}^2|$
of the covariance matrix. 
When homodyning the field, the random variable is $z\equiv
2e^{i\phi}x$ \cite{notaphi1} and the average over-line denotes the double 
integral over $x$ and $\phi$ in Eq. (\ref{sampO}). From Table (\ref{td1}) one has 
$\overline{z}=\<a\>$, $\overline{z^2}=\<a^2\>$, 
$\overline{|z|^2}=2\<a^{\dag}a\>+1$, 
$\overline{e^{2i\phi}}=0$ \cite{notaphi2}.
In this way one finds that the noise from homodyning the field is 
$N^{(\pm)}_{hom}[a]=1+2\<a^{\dag}a\>-|\<a\>|^2\pm|\<a^2\>-
\<a\>^2|$. On the other hand, when heterodyning, $z$ becomes the
heterodyne output photocurrent, whence
$\overline{z}=\<a\>$, $\overline{z^2}=\<a^2\>$, 
$\overline{|z|^2}=\<a^{\dag}a\>+1$, and one has
$N^{(\pm)}_{het}[a]=1+\<a^{\dag}a\>-|\<a\>|^2\pm|\<a^2\>-
\<a\>^2|$, so that the tomographical noise is larger than the heterodyne
noise by a term equal to the average photon number, i. e.
\begin{eqnarray}
N^{(\pm)}_{hom}[a]=N^{(\pm)}_{het}[a]+\<a^{\dag}a\>\;.
\end{eqnarray}
Therefore, homodyning the field is always more noisy than heterodyning it.
On the other hand, for other field observables it may happen that
homodyne tomography is less noisy than heterodyne detection. For
example, one can easily evaluate the noise $N_{hom}[\hat n]$ 
when homodyning the photon number $\hat n=a^{\dag}a$. The random 
variable corresponding to the photon number is
$\nu(z)={1\over2}(|z|^2-1)\equiv 2x^2-{1\over2}$, 
and from Table \ref{td1} we see that the noise 
$N_{hom}[\hat n]\doteq\overline{\Delta \nu^2(z)}$  can be written as 
$N_{hom}[\hat n]=\<\Delta\hat n^2\>+{1\over2}\<\hat n^2+\hat n+1\>$ \cite{nico}.
When heterodyning the field, the random variable corresponding to the
photon number is  $\nu(z)=|z|^2-1$, and from the relation 
$\overline{|z|^4}=\<a^{\dag}{}^2a^2\>$ one obtains 
$N_{het}[\hat n]\doteq\overline{\Delta \nu^2(z)}=
\<\Delta\hat n^2\>+\<\hat n+1\>$, namely
\begin{eqnarray}
N_{hom}[\hat n]=N_{het}[\hat n]+{1\over2}\<\hat n^2-\hat n-1\>\;.
\end{eqnarray}
We thus conclude that homodyning the photon number is less noisy
than heterodyning it for sufficiently low mean photon number $\<\hat n\><
{1\over2}(1+\sqrt{5})$.
\section{Homodyne tomography in presence of additive
phase-insensitive noise}
In this section I consider the case of additive Gaussian
noise, in the typical situation in which the noise is phase-insensitive.
This kind of noise is described by a density matrix evolved by the master equation
\begin{eqnarray}
\partial_t\hat\varrho(t)=2\left[ AL[a^{\dag}]+BL[a]\right]\hat\varrho(t)\;,\label{mm1}
\end{eqnarray}
where $L[\hat c]$ denotes the Lindblad super-operator $L[\hat c]\hat\varrho\doteq
\hat c\hat\varrho \hat c^{\dagger}-{1\over2}[\hat c^{\dagger}\hat
c,\hat\varrho]_+$. Due to the phase invariance $L[ae^{-i\phi}]=L[a]$
the dynamical evolution does not depend on the phase, 
and the noise is phase insensitive.
From the evolution of the averaged field $\langle a\rangle_{out}
\equiv\mbox{Tr}[a\hat\varrho(t)]
=g\langle a\rangle_{in}\equiv\mbox{Tr}[a\hat\varrho(0)]$ 
with $g=\exp[(A-B)t]$, we can see that for
$A>B$ Eq. (\ref{mm1}) describes phase-insensitive amplification with field-gain $g$,
whereas for $B>A$ it describes phase-insensitive attenuation, with $g<1$. 
Concretely, for $A>B$ Eq. (\ref{mm1}) models unsaturated parametric amplification with
thermal idler [average photon number $\bar m=B/(A-B)$], or unsaturated laser action 
[$A$ and $B$ proportional to atomic populations on the upper and lower 
lasing levels respectively]. For $B>A$, on the other hand,
the same equation describes a field mode damped
toward the thermal distribution [inverse photon lifetime
$\Gamma=2(B-A)$, equilibrium photon number $\bar m=A/(B-A)$], or a
loss $g<1$ along an optical fiber or at 
a beam-splitter, or even due to frequency 
conversion\cite{macchiavello}. The borderline case $A=B$ leaves the
average field invariant, but introduces noise that changes the
average photon number as $\langle a^{\dag}a\rangle_{out}=
\langle a^{\dag}a\rangle_{in}+\bar n$, where 
$\bar n=2At$. In this case the solution of Eq. (\ref{mm1})
can be cast into the simple form
\begin{eqnarray}
\hat\varrho(t)=\int\frac{d^2\beta}{\pi\bar{n}}\exp
\left(-|\beta|^2/\bar{n}\right)\hat D(\beta)\hat\varrho(0)\hat D^{\dag}(\beta)\;.
\end{eqnarray}
This is the {\em Gaussian displacement noise} studied in 
Refs.\cite{Hall,Hallpra} and commonly 
referred to as ``thermal noise'' [regarding the 
misuse of this terminology, see Ref. \cite{Hallpra}], which can be
used to model many kinds of undesired environmental effects, 
typically due to linear interactions with random classical fluctuating fields.
\par Eq. (\ref{mm1}) has the following simple Fokker-Planck differential 
representation \cite{nottingham} in terms of the generalized Wigner function 
$W_s (\alpha,\bar{\alpha})$ for ordering parameter $s$
\begin{eqnarray}
\partial_t 
W_s(\alpha,\bar{\alpha};t)=\left[ Q(\partial_{\alpha}\alpha+\partial_{\bar{\alpha}}
\bar{\alpha})+ 2D_s\partial^2_{\alpha,\bar{\alpha}}\right]W_s(\alpha,\bar{\alpha};t)
\;,\label{fp1}
\end{eqnarray}
where $Q=B-A$ and $2D_s=A+B+s(A-B)$. For nonunit quantum
efficiency $\eta$ and after a noise-diffusion time $t$
the homodyne probability distribution $p_{\eta}(x;\phi;t)$ 
can be evaluated as the marginal distribution of the Wigner 
function for ordering parameter $s={1-\eta^{-1}}$, namely
\begin{eqnarray}
p_{\eta}(x;\phi;t) =\int_{-\infty}^{+\infty}dy W_{1-\eta^{-1}}
\left((x+iy)e^{i\phi},(x-iy)e^{-i\phi};t\right)\;.\label{marg}
\end{eqnarray}
The solution of Eq. (\ref{fp1}) is the Gaussian convolution \cite{nottingham}
\begin{eqnarray}
W_s(\alpha,\bar{\alpha};t)&=&\int\frac{d^2\beta}{\pi\delta_s^2}
\exp\left[ -\frac{|\alpha-g\beta|^2}{\delta_s^2}\right]
W_s(\beta,\bar{\beta};0)\;,\qquad
\delta_s^2=\frac{D_s}{Q}(1-e^{-2Qt})\;,\label{fp2}
\end{eqnarray}
and using Eq. (\ref{marg}) one obtains the homodyne probability distribution
\begin{eqnarray}
p_{\eta}(x;\phi;t) =e^{Qt}\int_{-\infty}^{\infty}\frac{dx'}{\sqrt{2\pi
\Delta^2_{1-\eta^{-1}}}}
\exp\left[-\frac{(x'-g^{-1}x)^2}{2\Delta^2_{1-\eta^{-1}}}\right]p_{\eta}(x';\phi)\;.
\end{eqnarray}
where
$\Delta^2_{\eta}={1\over2}g^{-2}\delta^2_{1-\eta^{-1}}$. 
It is easy to see that the generating function of the 
$\hat{x}_{\phi}$-moments with the experimental probability
$p_{\eta}(x;\phi;t)$ can be written in term of the probability 
distribution $p(x;\phi)$ for perfect homodyning as follows
\begin{equation}
\int^{+\infty}_{-\infty} \! d x\, p_{\eta} (x;\phi;t)e^{ik
x} = \exp\left( -{1\over2}g^2\Delta^2_{\eta}k^2-
\frac{1-\eta}{8\eta}g^2k^2\right) \, 
\int^{+\infty}_{-\infty}\! d x\, p(x;\phi)e^{igk x} \;,\label{gc1}
\end{equation}
Eq. (\ref{gc1}) 
has the same form of Eq. (\ref{gc}), but with the Fourier
variable $k$ multiplied by $g$ and with an overall {\em effective quantum efficiency}
$\eta_*$ given by
\begin{eqnarray}
\eta_*^{-1}=\eta^{-1}+4\Delta^2_{\eta}=g^{-2}\eta^{-1}+\frac{2A}{B-A}(g^{-2}-1)
\;.\label{eta*}
\end{eqnarray}
On the other hand, following the same lines that lead us to Eq. (\ref{p2}),
we obtain the operator identity
\begin{equation}
\hat{\varrho}\equiv\hat{\varrho}(0) = \int^{\pi}_0 \frac{d\phi}{\pi} \,
\int^{+\infty}_{-\infty} \, d x\, p_{\eta_*} (x;\phi;t)
K_{\eta_*} (g^{-1}x-\hat{x}_{\phi}) \, , \label{p22}
\end{equation}
which also means that when homodyning the operator $\hat O$ one should
use ${\cal R}_{\eta_*}(g^{-1}x;\phi)$ in place of ${\cal R}_{\eta}(x;\phi)$,
namely, more generally, one needs to re-scale the homodyne outcomes by the gain and
use the effective quantum efficiency $\eta_*$ in Eq. (\ref{eta*}).
In terms of the gain $g$ and of the input-output photon numbers,
the effective quantum efficiency reads
\begin{eqnarray}
\eta_*^{-1}=\eta^{-1}+g^{-2}(2\langle a^{\dag}a\rangle_{out}+\eta^{-1})
-(2\langle a^{\dag}a\rangle_{in}+\eta^{-1})\;.\label{enn}
\end{eqnarray}
In the case of pure displacement Gaussian noise ($A=B$), Eq. (\ref{enn})  becomes 
\begin{eqnarray}
\eta_*^{-1}=\eta^{-1}+2\bar{n}\;,\label{oneph}
\end{eqnarray}
which means that the bound $\eta_* >1/2$ is surpassed already for
$\bar{n}\ge 1$: in other worlds, it is just sufficient to
have half photon of thermal noise to completely spoil the tomographic
reconstruction.
\section*{References}
\begin{description}
\itemsep=0pt\listparindent=0pt
\bibitem[1]{Raymer1} D. T. Smithey, M. Beck, M. G. Raymer, and
A. Faridani, Phys. Rev. Lett. {\bf 70}, 1244 (1993).
\bibitem[2]{our} G. M. D'Ariano, C. Macchiavello and M. G. A.
Paris, Phys. Rev. A{\bf 50}, 4298 (1994); Phys. Lett. A {\bf 195}, 31 (1994).
\bibitem[3]{munroe1} M. Munroe, D. Boggavarapu, M. E. Anderson, and
M. G. Raymer, Phys. Rev. A {\bf 52}, R924 (1995).
\bibitem[4]{schiller} S. Schiller, G. Breitenbach, S. F. Pereira, 
T. M\H{u}ller, and J. Mlynek, Phys. Rev. Lett. {\bf 77} 2933 (1996);
see also: G. Breitenbach, S. Schiller, and J. Mlynek, {\em Quantum
state reconstruction of coherent light and squeezed light} on this
volume.
\bibitem[5]{dlp} G. M. D'Ariano, U. Leonhardt and H. Paul,
Phys. Rev. A {\bf 52} R1801 (1995).
\bibitem[6]{dariJEOS} G. M. D'Ariano, Quantum Semiclass. Opt. {\bf 7}, 693 (1995).
\bibitem[7]{Richter} Th. Richter, Phys. Lett. A {\bf 221} 327 (1996).
\bibitem[8]{factulf} U. Leonhardt, M. Munroe, T. Kiss, Th. Richter,
and M. G. Raymer, Opt. Comm. {\bf 127}, 144 (1996).
\bibitem[9]{lpd} U. Leonhardt, H. Paul and G. M. D'Ariano,
Phys. Rev. A {\bf 52} 4899 (1995); H. Paul, U. Leonhardt, and G. M. D'Ariano, 
Acta Phys. Slov. {\bf 45}, 261 (1995).
\bibitem[10]{Bilk-poms} G. M. D'Ariano, {\em Quantum Estimation Theory
and Optical Detection}, in  {\em Concepts and Advances in Quantum Optics and 
Spectroscopy of Solids}, ed. by T. Hakioglu  and A. S. Shumovsky. 
(Kluwer, Amsterdam 1996, in press).
\bibitem[11]{Bilk-tomo} G. M. D'Ariano, {\em Measuring Quantum
States}, in the same book of Ref. (\cite{Bilk-poms}).
\bibitem[12]{single} G. M. D'Ariano and H. P. Yuen, Phys. Rev. Lett. 
{\bf 76} 2832 (1996).
\bibitem[13]{nico} G. M. D'Ariano, C. Macchiavello, and N. Sterpi,
{\em Systematic and statistical errors in homodyne measurements 
of the density matrix}, submitted to Phys. Rev. A.
\bibitem[14]{gradshtein} I. S. Gradshteyn and I. M. Ryzhik , {\em Table of
integrals, series, and products}(Academic Press, 1980).
\bibitem[15]{Rich} Th. Richter, Phys. Rev. A {\bf 53} 1197 (1996).
\bibitem[16]{baltin} R. Baltin, J. Phys. A Math. Gen. {\bf 16} 2721 (1983);
Phys. Lett. {\bf 102}A 332 (1984).
\bibitem[17]{yuen} H. P. Yuen, Phys. Lett. {\bf 91A}, 101 (1982).
\bibitem[18]{notaphi1} Notice that for the complex random variable $z=2
e^{i\phi}x$ the phase $\phi$ is a scanning parameter imposed by the detector.
(Actually, the best way to experimentally scan the integral in Eq. (\ref{sampO}) 
is just to pick up the phase $\phi$ at random.)
Nevertheless, the argument of the complex number $z$ is still a genuine
random variable, because the sign of $x$ is random, and depends on the value
of $\phi$. One has arg$(z)=\phi+\pi(1-\mbox{sgn}(x))$. For example,
for any highly excited coherent state $|\alpha\>$ the probability distribution of 
arg$(z)$ will approach a uniform distribution on $[\mbox{arg}(\alpha)-\pi/2,
\mbox{arg}(\alpha)+\pi/2]$.
\bibitem[19]{notaphi2} One should remember that, the phase $\phi$ is
imposed by the detector, and is uniformly scanned (randomly or not) 
in the interval $[0,\pi]$. This leads to $\overline{e^{2i\phi}}=0$,
independently on the state $\hat\varrho$. 
\bibitem[20]{macchiavello} G. M. D'Ariano and C. Macchiavello, Phys. Rev. {\bf A} 
{\bf 48} 3947, (1993).
\bibitem[21]{Hall} M. J. W. Hall, {\em Phase and noise} in Quantum
Communication and Measurement, ed. V. P. Belavkin, O. Hirota
and R. L. Hudson, Plenum Press (New York and London 1995), p. 53-59.
\bibitem[22]{Hallpra} M. J. W. Hall,  Phys. Rev. A {\bf 50} 3295 (1994).
\bibitem[23]{nottingham} G. M. D'Ariano, C. Macchiavello, and M. G. A. Paris,
{\em Information gain in quantum communication channels}, in
{\em Quantum Communication and Measurement}, ed. V. P. Belavkin, O. Hirota
and R. L. Hudson, Plenum Press (New York and London 1995), pag. 339.
\end{description}
\end{document}